\documentclass [12pt, peerreview]{IEEEtran}
\usepackage{color}
\usepackage{authblk}
\usepackage{algorithm}
\usepackage{algorithmic}
\usepackage{graphicx,amsmath,amssymb,latexsym,epsfig}
\usepackage{setspace}
\usepackage{enumerate}
\newtheorem{theorem}{Theorem}
\newtheorem{problem}{Problem}
\newtheorem{corollary}{Corollary}
\newtheorem{definition}{Definition}
\newtheorem{lemma}{Lemma}

\newcommand{\enp} {\hfill \rule{2.2mm}{2.6mm}}

\begin{document}

\def\eg{{e.g.}}
\def\ie{{i.e.}}
\bibliographystyle{IEEE}
\doublespacing

\title{Optimal Offline Broadcast Scheduling with an Energy Harvesting Transmitter}
\author{Hakan Erkal}
\author{F. Mehmet Ozcelik}
\author{Elif Uysal-Biyikoglu{\thanks{This work was partly supported by TUBITAK under grant 110E252. A preliminary version of these results was presented at ISIT 2011 St. Petersburg, Russia, Aug 2011~\cite{OzErU}.}}}
\affil{Dept. of Electrical and Electronics Eng., METU, Ankara 06531 Turkey\\
{herkal@ieee.org, mehmet.ozcelik@metu.edu.tr, elif@eee.metu.edu.tr}}
\bibliographystyle{IEEE}
\maketitle

\begin{abstract}
We consider an energy harvesting transmitter broadcasting data to two receivers. Energy and data arrivals are assumed to occur at arbitrary but known instants. The goal is to minimize the total transmission time of the packets arriving within a certain time window, using the energy that becomes available during this time.  An achievable rate region with structural properties satisfied by the two-user AWGN BC capacity region is assumed. Structural properties of power and rate allocation in an optimal policy are established, as well as the uniqueness of the optimal policy under the condition that all the data of the ``weaker" user are available at the beginning. An iterative algorithm, DuOpt, based on block coordinate descent that achieves the same structural properties as the optimal is described. Investigating the ways to have the optimal schedule of two consecutive epochs in terms of energy efficiency and minimum transmission duration, it has been shown that DuOpt achieves best performance under the same special condition of uniqueness. 
\end{abstract}


\newpage

\section{Introduction}
\label{sec:intro}

The basic offline problem of energy-efficient packet transmission  
scheduling ~\cite{UEP02, BeGa02,NuSr02,ZaMo09} is to assign code rates  
(consequently transmission durations) to a set of packets whose  
arrival times are known beforehand, so that they are all transmitted  
within a given time window with minimum total energy. The solution  
needs to strike a tradeoff between energy and delay based on the  
observation that energy per bit with many ideal and suboptimal coding  
schemes is convex and monotone increasing with rate. Recently, the  
problem has been reformulated with a model where energy gets  ``harvested" or replenished at certain known instants~\cite{YaU2010}.

While in the former formulations transmission rate needs to be adapted to the arrival rate of information here it is adapted to the generation rate of energy. Considering both of theses effects introduces a richness to the problem on top of the initial model. The point-to-point problem in~\cite{YaU2010} was recast for finite energy storage~\cite{TuYe2010} and for a wireless fading channel~\cite{OzTu2010}. 
The formulation has been extended to an AWGN BC
in~\cite{MAAEUHE2010,YaOU2010}, considering a static pool of data to be  
sent at the beginning of the schedule. The same BC problem was also studied under a limited battery constraint~\cite{OzYa2010}. 

The problem in~\cite{MAAEUHE2010} and \cite{YaOU2010} is reformulated in~\cite{OzErU2011} relaxing the assumption that data is ready at the beginning of the schedule. This paper, extending the work in~\cite{OzErU2011}, considers the broadcast problem where, given an average transmit power constraint, rates are picked from an achievable rate region which obeys certain structural properties satisfied by the AWGN BC. The sender (transmitter) gets replenished with arbitrary amounts of energy as well as data packets of arbitrary length destined to each user at arbitrary points in time.

The choices of power level and the rates to individual receivers across time is called a \emph{schedule}. An optimal scheduling policy is defined to be one that transmits all the bits that have arrived within a certain time window, in the minimum possible amount of time $T^{\rm{opt}}$. The policy is allowed to use as many energy harvests as it needs, provided it respects causality (no energy is used before it is harvested.) The problem considered in this paper is an \emph{offline} problem, where data arrival and energy harvest instants and amounts are assumed to be known in advance. Although this kind of prior information of data and energy arrivals may not be a widely applicable assumption to real-world problems, results obtained from this work help us understand the nature of an optimal solution and boundaries on the best performance.  Online formulations have also appeared in the literature. Notably, ~\cite{Shroff2011,Tassiulas2010} develop online scheduling policies for multihop networks on finite-horizon and infinite horizon problem formulations, respectively. 

To minimize the overall transmission duration, rates to the individual users should be chosen in a way such that transmission is fast and energy efficiency is satisfied. However, as we choose higher rates, we loose from energy efficiency. Balancing between fast and energy efficient transmission, the decision of rates needs to be based on the sequence of energy harvests and data arrivals. 

This paper essentially bridges the work that considered scheduling on a Broadcast Channel \emph{data} that becomes available at arbitrary points in time~\cite{UEP02} and work that considered chunks of \emph{energy} becoming available at arbitrary points in time~\cite{MAAEUHE2010}. It can also be viewed as the multiuser correspondent of the second problem considered in~\cite{YaU2010}. The main contributions are: (1) to establish structural properties of the optimal schedule, (2) the uniqueness of the optimal policy when all weaker user data is ready at the beginning, (3) an iterative algorithm which returns the optimal schedule under the same condition. It is shown in Section~\ref{sec:sm} that in an optimal policy, transmit power used is constant within each epoch, and may only rise from one epoch to the next, so that once it starts, the transmitter never lowers its power until it finally goes silent. On the other hand, the transmitter should increase its power only under certain conditions. These conditions, along with other structural properties of power and rate are established in Section~\ref{sec:structure}. Next, the uniqueness of the optimal policy is established under the condition that all of the weaker user's bits are available at the beginning. Finally, an iterative algorithm (that we refer to as DuOpt) based on the nonlinear block descent method which returns a feasible schedule carrying the same structural properties that the optimal is shown to have, is described. It has been shown that DuOpt returns the optimal schedule in case of static pool of weaker user data at the beginning of the schedule. We start by giving the problem statement in the next section.

\section{System Model}
\label{sec:sm}

\begin{figure}[hbt]
\centering \includegraphics[scale=0.5]{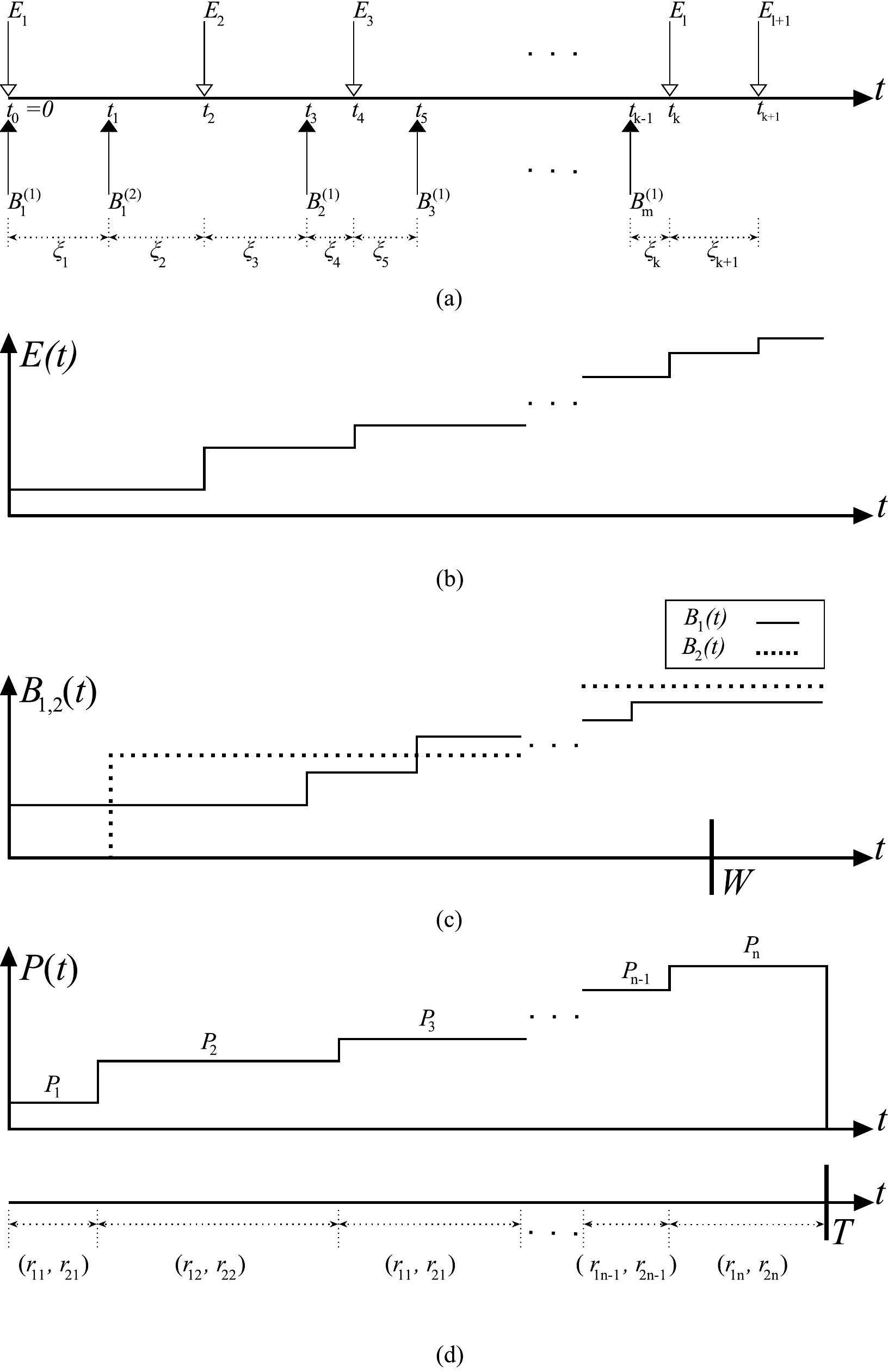}
\caption{Example for (a) a sequence of energy and data arrivals, (b)-(c) the
corresponding $E(t)$,$B(t)$, (d) the schedule $P(t)$ and $\{r_{1i},r_{2i}\}$.}
\label{fig:System_Model}
\end{figure}

Consider a broadcast channel with one transmitter and two receivers. Arbitrary amounts of energy, $\{E_i<\infty, i=1,2,\ldots\}$, as well as data for each user $\{B_i^{(1)},B_i^{(2)}<\infty, i=1,2,\ldots\}$ become available to the sender at arbitrary times $t_i$. A possible sequence of data and energy arrivals is illustrated in Fig.~\ref{fig:System_Model}. $E(t)$ denotes the total energy that has been {\emph{harvested}} in $[0,t)$ (regardless of how much of it has been used.) Similarly, $B_1(t)$ and $B_2(t)$ denote the total number of bits destined to the first and second user, respectively, that the sender has obtained in $[0,t)$. The interval between any two sequential arrival events (regardless of energy or data) will be called an inter-arrival {\emph{epoch}}. The length of the $i^{th}$ epoch is $\xi_i=t_{i}-t_{i-1}$. 

In this offline problem, all the future arrival times and amounts of energy and bits are known by the sender at $t=0$. It is also assumed that harvested energy and data are available for use instantaneously as they arrive, and code rate and transmission power decisions can be changed instantaneously. However, codeword block lengths will be chosen such that each codeword is sent completely within a single epoch (note that starting and ending times of epochs are known ahead of time), so that no arrival event occurs during a codeword. Consequently, the power and rate pair decision will be fixed throughout each codeword.


We are interested in minimizing the total transmission time for packets arriving by a certain time $W<\infty$, so W.L.O.G., set $B_i(t)=B_i(W)$ for $t > W$, $i=1,2$. A schedule, which is a sequence of power and rate allocations, is feasible if it sends $B_1(W)<\infty$ bits to the $1^{st}$ user and $B_2(W)<\infty$ to the $2^{nd}$ user (with a certain level of reliability\footnote{The achievable rate regions will be implicitly assumed to correspond to a certain constant tolerable error probability respecting which it is possible to transmit a finite number of bits with a finite amount of energy per bit.}), without violating causality (at any time, using available energy and data by that time). We are interested in finding among all feasible schedules one with the smallest completion time, $T^{\rm{opt}}$.

The structure of the achievable rate region will be based on the two-user AWGN BC. 
The capacity region of a two-user discrete time AWGN BC with average power constraint $P$, noise variance $\sigma^2$, where the $1^{st}$ user's channel gain ($s_1>0$)  is larger than the $2^{nd}$ user's ($s_2>0$), consists of rate pairs $(r_1, r_2)$ satisfying:
\begin{equation}
\small
\label{Eq:Broadcast_Channel}
r_1 \leq \frac{1}{2}\log_2\left(1+\frac{\alpha s_1 P}{\sigma^2}\right)\mbox{~,~} r_2 \leq \frac{1}{2}\log_2\left(1+\frac{(1-\alpha) s_2 P}{\alpha  s_2 P + \sigma^2}\right)
\end{equation} 
\normalsize
where $\alpha$, ($0 \leq \alpha \leq 1$), denotes the fraction of $P$ used for the $1^{st}$ user. Since $s_1>s_2$, the $1^{st}$  user will be referred as the ``stronger user", and the $2^{nd}$ as the ``weaker user". From (\ref{Eq:Broadcast_Channel}) each user's rate can be expressed as a function of the other's and $P$, as $r_1 = h_1(P,r_2)$, $r_2 = h_2(P,r_1)$. The rate functions $h_1$ and $h_2$ defined on $\Re^+\times \Re^+$ will be assumed to satisfy the following properties:
\begin{enumerate}
\item Nonnegativity: $h_1(P,r) \geq 0 , h_2(P,r)\geq 0$.
\item Monotonicity: $h_1(P,r)$, $h_2(P,r)$ are both monotone decreasing in $r$, and monotone increasing in $P$.
\item Concavity: $h_1(P,r)$, $h_2(P,r)$ are concave in $P$ and $r$:
$\frac{\partial^2 h_i(P,r)}{\partial P^2} \leq 0$,$\frac{\partial^2 h_i(P,r)}{\partial r^2} \leq 0$, for $i \in {1,2}$.
\item $\frac{\partial^2 h_1(P,r)}{\partial r \partial P}\geq 0$, $\frac{\partial^2 h_1(P,r)}{\partial P \partial r } \leq 0.$
\item $\frac{\partial^2 h_2(P,r)}{\partial r \partial P}=0$, $\frac{\partial^2 h_2(P,r)}{\partial P \partial r }=0.$
\end{enumerate}

The results in the rest of the paper will be valid for any rate function satisfying (1)-(5), which are also satisfied by the AWGN BC~\cite{MAAEUHE2010}.

It is straightforward to show that one can restrict attention to feasible schedules that do not change their power and rate allocations within epochs.

\begin{lemma}
\label{lmm:ConstantPowerRate}
In an optimal schedule, the power and rate pair remain constant within all epochs, except for the epoch during which the schedule ends.
\end{lemma} 
\noindent{\emph{Proof.}} During an epoch, there are no energy or data arrivals and the claim is identical with the one stated and proved in Lemma 2 of~\cite{MAAEUHE2010}. The power will drop to zero when the schedule ends, which is in general within (and not necessarily at the end of) the last epoch used by the schedule. \enp

With this, we will take rate and power assignments constant during an epoch. Let $P_i$ be the total transmit power and $r_{ji}$ be the rate assigned to the $j^{th}$ user during epoch $i$. Similarly, $P_{ji}$ represent the power assigned to $j^{th}$ user during epoch $i$. We are now ready to state the problem in terms of power and rate allocations to epochs, more precisely, an assignment of power and the stronger user's rate to each epoch (the weaker user's rate is thus determined). A final technical assumption will be useful in stating the problem: we shall assume that there is some $k^{\rm up}<\infty$ such that there is at least one feasible schedule that ends within the first $k^{\rm up}$ epochs. In other words, $k^{\rm up}$ is an upper bound for epochs to be considered. In problem statement, $k^*$ denotes the last epoch of an optimal schedule, where $k^* \leq k^{\rm up}$.\\

\begin{problem}
\label{pr:MultiuserScheduling}
\noindent{\bf Transmission Time Minimization of Data Arriving at Arbitrary Points on an Energy Harvesting BC:}
\small \begin{align}
\noindent \mbox{Minimize:  } &T=T(\{P_i,r_{1i}\}_{1 \leq i \leq k^{\rm up}}) \nonumber \\
\noindent \mbox{subject to: }&P_i\geq 0 \mbox{~,~}0 \leq r_{1i} \leq h_1(P_i,0)\mbox{~,~}r_{2i}=h_2(P_i,r_{1i})\nonumber\\
&\sum_{i=1}^{k} P_i\xi_{i} \leq E(t_k) \nonumber\\
&\sum_{i=1}^{k^{*}}P_i \xi_i+P_{(k^{*}+1)}(T-\sum_{i=1}^{k^{*}}\xi_{i})\leq E(T) \label{eq:EConst1}\\
&\sum_{i=1}^{k} r_{1i}\xi_{i} \leq B_1({t_k}) \mbox{~,~} \sum_{i=1}^{k} r_{2i}\xi_{i} \leq B_2(t_k) \label{eq:B1B2Const1}\\
&for \mbox{ }k=1,2,...,k^{*}=\max\{i:\sum_{j=1}^{i} \xi_j < T\}\nonumber\\
&\sum_{i=1}^{k^{*}} r_{1i}\xi_{i}+r_{1(k^{*}+1)}(T-\sum_{i=1}^{k^{*}}\xi_{i})=B_1(T) \nonumber\\
&\sum_{i=1}^{k^{*}} r_{2i}\xi_{i}+r_{2(k^{*}+1)}(T-\sum_{i=1}^{k^{*}}\xi_{i})=B_2(T)\label{eq:B1B2Const2}
\end{align} \normalsize
\end{problem}

We will refer to~\eqref{eq:EConst1} and~\eqref{eq:B1B2Const1} as energy and data causality constraints, respectively, as these ensure no energy is consumed and no bit is transmitted before becoming available. In addition, when the $k^{th}$ inequality in~\eqref{eq:EConst1} holds with equality, we shall say that $k^{th}$ \emph{energy constraint is active}. Similarly, equality case in~\eqref{eq:B1B2Const1} will be referred as a \emph{data constraint being active}. Finally, the feasibility constraint \eqref{eq:B1B2Const2}, ensures all the data bits destined to each user are transmitted. 

In the next section, we investigate structural properties that any optimal schedule has to satisfy.

\section{Structure of an Optimal Policy}
\label{sec:structure}
Lemma~\ref{lmm:ConstantPowerRate} recorded that in an optimal schedule power can only change upon a data arrival or energy harvest. The next result states that when power changes, it can only increase. The key to the proof is that more ``bits per joule" can be sent by evenly distributing energy across a time interval (\ie, maintaining a constant power level, which is a consequence of the convexity properties of our rate functions.) If an even distribution of power requires transferring energy or bits to the latter epoch, it can always be done; hence, total transmit power never decreases in time. But, power may increase in time, because even distribution of power may result in unmet causality constraints. We state these results in Lemma~\ref{lmm:PowernonDecreasing}.

Due to space constraints, the proofs of the following results (Lemma~\ref{lmm:PowernonDecreasing} through Lemma~\ref{lmm:ConstantRate1}) are omitted, and given in~\cite{Hakan_Thesis}.

\begin{lemma}
\label{lmm:PowernonDecreasing}
\emph{(For proof see~\cite{Hakan_Thesis})} Consider an optimal schedule that ends during epoch $k^*$. Power is non-decreasing with epoch index, i.e, $P_i\leq P_{i+1}$ for $i=1,2,\ldots k^*-1$.
\end{lemma}

As stated in Lemma~\ref{lmm:PowernonDecreasing} power cannot decrease, yet may rise in time. In the next Lemma we note what is necessary condition for such a rise to occur in an optimal policy.

\begin{lemma}
\label{lmm:CloserPowers}
\emph{(For proof see~\cite{Hakan_Thesis})} In an optimal policy, power can only rise at $t_i$ (end of epoch $i$) if at least one of the conditions below holds:
\renewcommand{\theenumi}{\alph{enumi}}
\begin{enumerate}

\item \emph{Energy constraint is active at point $t_i$. (\ie the $i{th}$ energy constraint is active)}  \label{cnd:a} 
\item \emph{The data constraints for both users are active at point $t_i$. (\ie, the set of constraints in~\eqref{eq:B1B2Const1})} \label{cnd:b}
\item \emph{The weaker user's data constraint is active and data arrival to the weaker user occurs at time $t_i$.}\label{cnd:c}

\end{enumerate}
\end{lemma}

The next set of results illustrate the structure of {\emph{rate allocation}} in conjunction with the power allocation in an optimal policy.

\begin{corollary}
\label{corollary:PowerIncrease}
\emph{(For proof see~\cite{Hakan_Thesis})} In an optimal policy,
\begin{enumerate}
\item If power increases upon a data arrival for the second user, data to be sent to the weaker user have been finished by this event.
\item If power rises upon a data arrival for the stronger user, all available bits have been sent by this event.
\item If power increases upon an energy harvest, all energy available at the beginning of the former constant power band has been consumed by this energy harvest.
\end{enumerate}
\end{corollary}

In the rest, some properties will be proved under the condition that all weaker user data is available at the beginning. We shall abbreviate this condition as follows:\\

\begin{definition}
{\emph{Weaker User Full Buffer Condition (WUFBC)}} is said to be satisfied whenever all of the data of the weaker user is available at the beginning of transmission. That is, $B_2(W)=B_2(0)$.
\end{definition}

The following lemma states an important feature of the stronger user rate distribution under WUFBC.

\begin{lemma}
\label{lmm:ConstantRate1}
\emph{(For proof see~\cite{Hakan_Thesis})} Consider two consecutive epochs $i$ and $i+1$ of a given schedule, ending at $t_{i}$ and   $t_{i+1}$ by definition, and suppose WUFBC holds for the problem instance. The following is necessary for the rate and power allocation to these two epochs of the given schedule to be locally optimal:  The stronger user's rate is constant throughout $[t_{i-1},t_{i})$, and $[t_{i},t_{i+1})$. Furthermore, the rate may jump up at  $t=t_{i}$ (staying constant otherwise) if at least one of the below is true:
\begin{enumerate}
\item There is data arrival to the stronger user at $t=t_{i}$ and all the data that arrived before $t=t_{i}$ has been transmitted by $t_{i}$.
\item An energy harvest occurs at $t=t_{i}$ and all of the power has been used for the stronger user during epoch $i$.
\end{enumerate}
\end{lemma}

We investigate the unique solution of Problem~\ref{pr:MultiuserScheduling} in the next section.

\section{Uniqueness of the Optimum Schedule Under WUFBC}

In the following lemma we note that an optimal schedule uses all energy harvested by the time the schedule ends completely.
\begin{lemma}
\label{lmm:Consumed_Energy}
The energy consumed by an optimal schedule that ends at $T^{\rm {opt}}$ is equal to $E(T^{\rm {opt}})$.
\end{lemma} 

\noindent{\emph{Proof.}} To reach contradiction, consider an optimal schedule that consumes less energy than it harvested and has leftover energy in its energy buffer at $T^{\rm {opt}}$. The remaining energy in the buffer could have been used in the last epoch to decrease the transmission completion time, which contradicts the minimality of $T^{\rm {opt}}$. Hence, this schedule cannot be optimal. \enp

Next, we show the uniqueness of the optimal schedule under WUFBC.
\begin{theorem}
\label{thm:Unique_Schedule}
\emph{There is a unique optimum schedule under WUFBC, \ie, a unique power-rate allocation achieving $T^{\rm {opt}}$.}
\end{theorem}

\noindent{\emph{Proof.}}
Suppose that there are two distinct optimal schedules, $S^A$ and $S^B$, which have equal power and rate assignments until $t_{s-1}$ and differ for the first time at epoch $s$. Consider that the corresponding power allocation vectors, $\textbf{P}^A$ and $\textbf{P}^B$, also differ at epoch $s$ such that $P_i^A = P_i^B ,\forall i \in \{1,2,..,s-1\}$ and $P_s^A<P_s^B$. First, assume that $\textbf{P}^A$ remains constant after epoch $s$, \ie, $P_i^A=P_s^A, \forall i>s$. By definition, both schedules end at $T^{\rm{opt}}$. The total energy consumption of $S^A$ would be less than that of $S^B$ by $T^{\rm{opt}}$, \ie, $\left(\sum_{i=1}^{k^*} P_i^A \xi_i + (T^{\rm{opt}} - t_{k^*+1}) P_{k^*+1}^A\right) <\left(\sum_{i=1}^{k^*} P_i^B \xi_i + (T^{\rm{opt}} - t_{k^*+1}) P_{k^*+1}^B\right)$ , which contradicts Lemma~\ref{lmm:Consumed_Energy}. Hence, total transmit power of $S^A$ cannot remain constant after $t_s$. Since total transmit power is nondecreasing (See Lemma~\ref{lmm:PowernonDecreasing}), it should increase after epoch $s$ and before the end of transmission, \ie, $P_{u}^A<P_{u+1}^A \mbox{~,~} \exists u \in \{s,s+1,...,k^*\}$. Since there are no data arrivals for the weaker user, the increase in total transmit power is either due to energy constraint being met or due to \emph{all} the packets arrived by the time $t_u$ having been transmitted (cf. conditions (a) or (c) in Lemma~\ref{lmm:CloserPowers}). As $\sum_{i=s}^{u}P_i^A \xi_i<\sum_{i=s}^{u}P_i^B \xi _i$, $S^A$ has not consumed all the available energy at the end of epoch $u$. Hence, $S^A$ must have transmitted all the bits arrived until $t_{u}$, which means that $S^A$ has transmitted at least the same number of bits to both users while consuming less energy than $S^B$ between $t_0$ and $t_{u}$, which contradicts the optimality of $S^B$. Therefore, if there are two distinct optimal schedules, $S^A$ and $S^B$, their power allocation vectors cannot be different, \ie, $\textbf{P}^A = \textbf{P}^B$.

Now, consider two rate pair vectors, $\textbf{R}^A$ and $\textbf{R}^B$, where $(r_{1i}^A,r_{2i}^A) = (r_{1i}^B,r_{2i}^B)\mbox{~,~} \forall i \in \{1,2,..,s-1\}$ and $r_{1s}^A < r_{1s}^B$. Let the rate of the stronger user in $S^A$, $\{r_{1j}^A\}$ stay constant after $t_{s-1}$. By Lemma~\ref{lmm:ConstantRate1} rate of the stronger user cannot decrease, hence the rate of the stronger user in $S^B$ would be larger than that of $S^A$ after epoch $s$, \ie, $r_{1(j+1)}^A = r_{1s}^A<r_{1s}^B\leq r_{1j}^B \mbox{~,~} \forall j \in \{s,s+1,...,k-1\}$. Since both schedules end transmission at the same time, $S^A$ transmits fewer bits to the stronger user than $S^B$ does, which contradicts the fact that optimal schedule transmits all the packet arrivals by the end of transmission. Therefore, the rate of the stronger user in $S^A$ cannot stay constant after epoch $s$. Now suppose that rate of the stronger user in $S^A$ increases at the end of epoch $u$, \ie, $r_{1u}^A < r_{1(u+1)}^A \mbox{~,~} \exists u \in \{s,s+1,...,k^*\}$. This increase cannot be due to (1) in Lemma~\ref{lmm:ConstantRate1} because $S^B$ has transmitted more bits to the stronger user by $t_{u}$, \ie, $\sum_{i=1}^{u}r_{1i}^A \xi_i < \sum_{i=1}^{u}r_{1i}^B \xi_i$. Moreover, this increase cannot be due to (2) in Lemma~\ref{lmm:ConstantRate1} since rate of the weaker user in $S^A$ is greater than zero in epoch $u$, \ie, $r_{2u}^A = h_2(P_u,r_{1u}^A) > h_2(P_u,r_{1u}^B)\geq 0$. Hence rate of stronger user in $S^A$ cannot increase after epoch $s$. Finally, rate of the stronger user in $S^A$ cannot {\emph{decrease}} (See Lemma~\ref{lmm:ConstantRate1}) as this would also contradict optimality. Hence, there cannot be two optimal schedules with different rate pair vectors. 

As both the power allocation vector and the rate pair vector of an optimal schedule are unique, we conclude that the optimal schedule is \emph{unique} under WUFBC. \enp

\section{The DuOpt Algorithm}

The problem in~\cite{MAAEUHE2010} which is a special case of Problem~\ref{pr:MultiuserScheduling}, where both users' data is available at the beginning, was shown to be solved in~\cite{MAAEUHE2010} by the \emph{FlowRight} algorithm~\cite{Eu04}. Along similar lines, we develop an algorithm that we call \emph{DuOpt} for solving Problem~\ref{pr:MultiuserScheduling} in its general form. As a matter of fact, \emph{DuOpt} simply reduces to \emph{FlowRight} when the given problem instance has all the data arriving at $t = 0$. Similarly to \emph{FlowRight}, \emph{DuOpt} starts with any feasible schedule and reduces the transmission completion time iteratively. Let the number of epochs and the transmission completion time of the initial schedule be $k^{up}$ and $T^{up}$ respectively. In each iteration, \emph{DuOpt} sequentially updates rates and powers of two consecutive epochs at a time, \ie, epochs $(1,2),(2,3),...$, until all epochs are updated. Then, starting from the first epoch pair, \emph{DuOpt} continues with the next iteration. \emph{DuOpt} stops after $N$ iterations such that $N=\min\{n: T^{n-1} - T^{n} \leq \epsilon,i=1,...,k^{n}, j=1,2\}$, where $T^{n} \leq T^{\rm up}$ is the transmission completion time, $k^{n}\leq k^{\rm up}$ is the number of epochs used at the end of $n^{th}$ iteration and $\epsilon$ is a predefined threshold.

Hereafter, we will briefly outline the local optimizations over epoch pairs. In Theorem~\ref{thm:duopt_stops}, it will be shown that local optimizations can only improve the schedule. We will also prove that under WUFBC, successive iterations strictly improves the schedule unless it is optimal.

\subsubsection*{Local Optimizations}
Let $E_i^n$ denote the energy used during the $i^{th}$ epoch and $b_{ji}^n$ denote the number of bits transmitted to the $j^{th}$ user during epoch $i$ at the end of $n^{th}$ iteration. Suppose that \emph{DuOpt} is at the $n^{th}$ iteration and running a local optimization over epoch pair $(i,i+1)$. The values of $b_{jz}^n$ and $E_z^n$, $\forall z \in \{1,2,...,i-1\}$ have already been found by previous local optimizations. At the end of this optimization, $E_i^n$ and $b_{ji}^n$ will be determined; $E_{i+1}^{n-1}$, $E_{i+2}^{n-1}$ and $b_{j(i+1)}^{n-1}$ will be reset to new values that conserve total energy consumption and data transmission in these epochs. The goal of the local optimization is surely to minimize the total transmission completion time of all the packet arrivals. Hence, it is logical to minimize the transmission time in the local optimization problem, which results in a gap\footnote{A gap is a time period with zero power allocation.} if transmission ends before the end of $(i+1)^{th}$ epoch. This gap is used in the next local optimization to further reduce the transmission time via transferring bits or energy between epochs $(i+1)$ and $(i+2)$; hence, a new gap occurs at the end of the next local optimization. This new gap propagates to the end of the transmission resulting a reduction in the total transmission completion time~\cite{MAAEUHE2010}. However, in some cases an epoch long gap occurs and this gap is useless for the next local optimization, \ie, energy or data transfer between epochs in the next local optimization is impossible because of constraints. In that case, it is better to just spread the data out till the end of the second epoch in the local problem and minimize the energy consumption so that the excess energy can be used to further reduce the transmission time in the next local optimization. This leads to two different local optimization functions: \emph{time minimization} and \emph{energy minimization}. These functions both support the global objective in different ways. Time minimization aims to find the minimum amount of time, $T_{(i,i+1)}^n$, to transmit $b_{j(i,i+1)}^n = b_{ji}^{n-1} + b_{j(i+1)}^{n-1}$ bits to each user using the energy available in epoch pair $(i,i+1)$, \ie, $E_{(i,i+1)}^{n-1}=E_i^{n-1}+E_{i+1}^{n-1}$. On the other hand, energy minimization aims to find the minimum energy, $E_{(i,i+1)}^n$, to transmit $b_{j(i,i+1)}^n$ bits to each user in two epoch durations, \ie, $\xi_i + \xi_{i+1}$, and excess energy, $ E_i^{n-1} + E_{i+1}^{n-1} - E_{(i,i+1)}^n$, is transferred to the $(i+2)^{th}$ epoch in order to conserve energy. Both of the optimizations respect energy and bit causalities, \ie, $E_{i}^n \leq E(t_i) - \sum_{s=1}^{i-1}E_{s}^n$ and $b_{ji}^n \leq B_j(t_i) - \sum_{s=1}^{i-1}b_{js}^n$, $ j \in \{1,2\}$. For details of the local optimization, see~\cite{Hakan_Thesis}.\\

\begin{figure}[htpb]
\begin{center}
\includegraphics[scale=0.5]{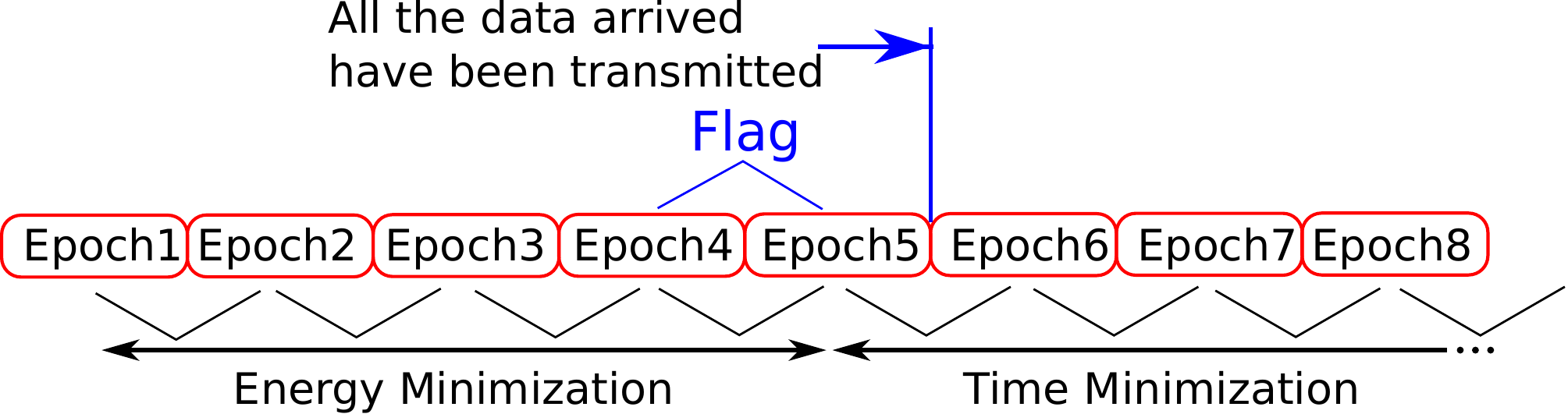}
\caption{Illustration of the \emph{Flag} and local optimizations, where all the feasible bits have been transmitted until the end of $5^{th}$ epoch; hence, a \emph{Flag} is set to $4^{th}$ epoch pair, \ie, $(4,5)$. Energy minimization is performed upto epoch pair with the \emph{Flag} and time minimization is performed for the rest.}
\label{fig:DuOptFlag}
\end{center}
\end{figure}

Suppose that all the feasible packets have been transmitted until the end of the $i^{th}$ epoch and there are still packets to arrive after $t_{i}$. Then, further minimization of transmission completion time of sequential epochs before $t_i$ will be suboptimal. On the other hand, we can minimize the energy consumption until $t_i$ and use the excess energy to minimize the transmission completion time. Therefore, utilization of energy minimization for local optimizations in Problem~\ref{pr:MultiuserScheduling} is very crucial. If it is guaranteed that current schedule uses at least the same amount of energy as optimal schedule until $t_i$, \emph{DuOpt} uses the energy minimization function upto $i^{th}$ epoch pair and the time minimization function for the rest. In order to determine when to switch from energy minimization to time minimization, a \emph{Flag} is placed at $i^{th}$ epoch pair. Initially, the \emph{Flag} is set to zero and \emph{DuOpt} starts with performing time minimization on epoch pairs. During $n^{th}$ iteration, if all the feasible bits are transmitted by the $i^{th}$ epoch for $\exists i \in \{1,2,...,k^{n}\}$, then the \emph{Flag} is set to $i$ $(Flag < i)$. In the following iterations, energy minimization function is used up to $i^{th}$ epoch pair. Fig.~\ref{fig:DuOptFlag} illustrates the \emph{Flag} usage and the pseudo-code in Algorithm~\ref{alg:DuOpt} outlines the DuOpt algorithm.\\


\begin{algorithm}
\caption{DuOpt Algorithm}
\label{alg:DuOpt}
\begin{algorithmic}[1]
	\STATE Initialize();
	\STATE {n $\gets$ 0, Flag $\gets$ 0, $T^{0} \gets T^{up}$}
	\REPEAT
	\STATE {n++} 
	\FOR {$i=1$ to $(k^{n}-1)$}
		\STATE $e_{i,max}^{n} \gets E(t_i)-\sum_{m=1}^{i-1} e_m^{n}$	
		\STATE $b_{1i,max}^{n} \gets B_1(t_i) -\sum_{m=1}^{i-1} b_{1m}^{n}$
		\STATE $b_{2i,max}^{n} \gets B_2(t_i) -\sum_{m=1}^{i-1} b_{2m}^{n}$
		\STATE $b_1^{n} \gets b_{1i}^{n-1} +  b_{1(i+1)}^{n-1}$
		\STATE $b_2^{n} \gets b_{2i}^{n-1} +  b_{2(i+1)}^{n-1}$
		\IF{$i \leq {\rm Flag}$}
		\STATE {[$b_{1i}^{n}$\verb+,+$b_{1(i+1)}^{n-1}$\verb+,+$b_{2i}^{n}$\verb+,+$b_{2(i+1)}^{n-1}$\verb+,+$E_i^n$\verb+,+$E_{i+1}^{n-1}$\verb+,+$E_{i+2}^{n-1}$] =\\ \verb+     + \verb+Minimize_Energy(+$E_i^{n-1}$\verb+,+$E_{i+1}^{n-1}$\verb+,+$E_{i+2}^{n-1}$\verb+,+$b_1^n$\verb+,+$b_2^n$\verb+,+$e_{i,max}^{n}$\verb+,+$b_{1i,max}^{n}$\verb+,+$b_{2i,max}^{n}$\verb+)+}
		\ELSE 
		\STATE {[$b_{1i}^{n}$\verb+,+$b_{1(i+1)}^{n-1}$\verb+,+$b_{2i}^{n}$\verb+,+$b_{2(i+1)}^{n-1}$\verb+,+$E_i^n$\verb+,+$E_{i+1}^{n-1}$] = \\ \verb+     + \verb+Minimize_Time(+$E_i^{n-1}$\verb+,+$E_{i+1}^{n-1}$\verb+,+$b_1^n$\verb+,+$b_2^n$\verb+,+$e_{i,max}^{n}$\verb+,+$b_{1i,max}^{n}$\verb+,+$b_{2i,max}^{n}$\verb+)+}
		\ENDIF
		\IF {$b_{1i,max}^{n} == b_{1i}^{n}~~\&\&~~b_{2i,max}^{n} == b_{2i}^{n}~~\&\&~~{\rm Flag} < i~~\&\&~~i<k^{n}-1$}
		\STATE {${\rm Flag} = i$}
		\ENDIF
	\ENDFOR
	\STATE \verb+Calculate_T(&+$T^n$ \verb+)+ \COMMENT{Calculate current transmission completion time.}
	\UNTIL {$T^n == T^{n-1}$}
	
\end{algorithmic}
\end{algorithm}

\begin{theorem}
\label{thm:duopt_stops}
\emph{Following statements hold:}
\begin{enumerate}
\item \emph{Successive iterations of DuOpt can only improve the schedule.}
\item \emph{DuOpt stops and returns a schedule with $\{r_{1i}^{\infty},r_{2i}^{\infty}\}$.}
\end{enumerate}
\end{theorem}

\noindent{\emph{Proof.}}
\begin{enumerate}
\item Suppose that DuOpt is running its $n^{th}$ iteration. After the local optimization on $i^{th}$ epoch pair, we obtain $\{(r_{1i}^n,r_{2i}^n) , E_i^n\}$ and reset the values of $\{(r_{1(i+1)}^{n-1},r_{2(i+1)}^{n-1}) , E_{i+1}^{n-1} , E_{i+2}^{n-1}\}$. If the \emph{Flag} is not placed before $i^{th}$ epoch pair, \ie, $Flag \geq i$, then the aim of the local optimization will be energy minimization. Following the local optimization on $i^{th}$ epoch pair, the excess energy will be transferred to $E_{i+2}^{n-1}$. In the next local optimization this excess energy is either further transferred or is used to reduce the transmit time. On the other hand, if $Flag < i$, then the aim of the local optimization on $i^{th}$ epoch pair will be time minimization. After the local optimization the transmission completion time of the bits in epochs $(i,i+1)$ will either be equal to or before the end of the epoch  $(i+1)$. That is, a gap may occur within $i^{th}$ epoch pair. In the next local optimization, this gap would propagate to the $(i+2)^{th}$ epoch~\cite{MAAEUHE2010}. During the $n^{th}$ iteration of \emph{DuOpt}, if a gap occurs or excess energy is transferred during local optimizations, then the gap (or the excess energy respectively) will propagate to the last epoch pair resulting in an ultimate reduction the transmission completion time at the end of the iteration, \ie, $T({r_{1i}^n,r_{2i}^n}) < T({r_{1i}^{n-1},r_{2i}^{n-1}})$. If neither excess energy nor a gap occurs during local optimizations, then transmission completion time cannot be decreased and \emph{DuOpt} will stop by definition.
 
Both local optimizations are in favor of the next local optimization. Therefore, if in either one of the local optimizations a gap occurs or excess energy is transferred then it would propagate till the last epoch pair and finally the transmission completion time would decrease at the end of $n^{ th}$ iteration, \ie, $T({r_{1i}^n,r_{2i}^n}) < T({r_{1i}^{n-1},r_{2i}^{n-1}})$. If neither excess energy nor gap occurs during local optimizations, then transmission completion time would not be decreased and \emph{DuOpt} would stop. \enp

\item In Part-1 we have shown that transmission completion time, $T({r_{1i}^n,r_{2i}^n})$, is strictly decreasing in each iteration; meanwhile it is bounded below by $T^{OPT}$. Therefore, the iterations of \emph{DuOpt} stop and return a schedule $\{r_{1i}^{\infty},r_{2i}^{\infty}\}$. \enp

\end{enumerate}

\subsubsection*{Optimality of the DuOpt algorithm under WUFBC}

\begin{theorem}
\label{thm:DuOpt_optimal}
\emph{If WUFBC is guaranteed, the schedule returned by \emph{DuOpt} is optimal, \ie,}
$T(\{r_{1i}^{\infty},r_{2i}^{\infty}\})=T^{\rm {opt}}$.
\end{theorem}

\noindent{\emph{Proof.}}
Suppose that \emph{DuOpt} stopped and returned a schedule $\{r_{1i}^{\infty},r_{2i}^{\infty}\} \triangleq S^{\rm Du}$, with completion time $T(\{r_{1i}^{\infty},r_{2i}^{\infty}\})\triangleq T^{\rm Du}$. Let $S^{\rm opt}$ be the unique optimal schedule with transmission completion time $T^{\rm opt}$. We will now prove that $S^{\rm Du} = S^{\rm opt}$. Let us suppose $S^{\rm Du} \neq S^{\rm opt}$, then these schedules have to differ in either the power allocation or rate allocation (or both). First, suppose $P_{i}^{\rm Du} = P_{i}^{\rm opt}, i \in \{1,2,...,s-1\}$ and $P_{s}^{\rm Du} \neq P_{s}^{\rm opt}$ for some $s$. We will show that this case is impossible. There are two possible cases for epoch $s$: (i) $P_{s}^{\rm Du} > P_{s}^{\rm opt}$, (ii) $P_{s}^{\rm Du} < P_{s}^{\rm opt}$. Let us begin with the first case.

\begin{enumerate}[(i)]
\item \label{prf:item1} We assumed $P_{s}^{\rm Du} > P_{s}^{\rm opt}$. If $P^{\rm opt}$ stays constant after epoch $s$ till the end of transmission, this would mean that $S^{\rm Du}$ consumes more energy than $S^{\rm opt}$ until  $T^{\rm opt}$, which would contradict the fact that optimal schedule consumes all the harvested energy till the end of transmission. Therefore the power of the optimal schedule must increase at the end of epoch $(s+m)$ for some $m \geq 0$ before the end of transmission. As $S^{\rm Du}$ has been able to use more energy than the optimal schedule until $t_{s+m}$, the optimal schedule cannot have run into an energy constraint at $t_{s+m}$, hence the rise in the power can only be due to a data constraint at $t_{s+m}$, \ie, all the bits arrived have been transmitted by the optimal schedule until $t_{s+m}$. In order to contradict the assumption that $P_{s}^{\rm Du} > P_{s}^{\rm opt}$, we will now analyze the rate assignments for both schedules. First let us focus on the case that both schedules use exactly the same rates for the stronger user up to $t_{s+m}$, \ie, $r_{1i}^{\rm Du} = r_{1i}^{\rm opt} \mbox{~,~} \forall i \in \{s,...,s+m\}$. As we have shown above, $S^{\rm opt}$ should have transmitted all the bits available until $t_{s+m}$. However, if we compare the weaker user bits transmitted by both schedules until $t_{s+m}$, we observe that $S^{\rm Du}$ transmits more bits to the weaker user than $S^{\rm opt}$ does, because $\sum_{i=1}^{s+m}(r_{2i}^{\rm Du} - r_{2i}^{\rm opt})\xi_i = \sum_{i=s}^{s+m}(r_{2i}^{\rm Du} - r_{2i}^{\rm opt})\xi_i = \sum_{i=s}^{s+m}(h_{2}(P_{i}^{\rm Du},r_{1i}^{\rm Du}) - h_{2}(P_{i}^{\rm opt},r_{1i}^{\rm opt}))\xi_i > 0$. On the other hand, \emph{DuOpt} respects \emph{bit causality}, \ie, \emph{DuOpt} does not transmit bits that have not arrived yet, so we reach contradiction. That is, rates cannot stay constant up to $t_{s+m}$, \ie, there is some $k \in \{1,2,...,s+m-1\}$ such that  $r_{1i}^{\rm Du} = r_{1i}^{\rm opt}$ for $i < k$ and $r_{1k}^{\rm Du} \neq r_{1k}^{\rm opt}$. But we shall now show that this is not possible. First consider the case that $r_{1k}^{\rm Du} < r_{1k}^{\rm opt}$. From Lemma~\ref{lmm:ConstantRate1},  the stronger user's rate cannot decrease under WUFBC. If $r_{1i}^{\rm Du} = r_{1k}^{\rm Du}, i \in \{k,...,s+m\}$, then $S^{\rm Du}$ transmits more bits to the weaker user than $S^{\rm opt}$ does by $t_{s+m}$, \ie, $\sum_{i=1}^{s+m}(r_{2i}^{\rm Du} - r_{2i}^{\rm opt})\xi_i = \sum_{i=k+1}^{s+m}(r_{2i}^{\rm Du} - r_{2i}^{\rm opt})\xi_i = \sum_{i=k+1}^{s+m}(h_{2}(P_{i}^{\rm Du},r_{1i}^{\rm Du}) - h_{2}(P_{i}^{\rm opt},r_{1i}^{\rm opt}))\xi_i > 0$. However, at the end of $(s+m)^{th}$ epoch, $S^{\rm Du}$ cannot send more bits to weaker user because $S^{\rm opt}$ should have transmitted all the weaker user bits. Therefore, $r_1^{\rm Du}$ should increase before $t_{s+m}$, \ie, at the end of epoch $k+n$, where $0 < n < (s+m-k)$. We have $r_{1(k+n)}^{\rm Du} < r_{1(k+n+1)}^{\rm Du}$, hence either one of the two conditions in Lemma~\ref{lmm:ConstantRate1} must hold. Since $\sum_{i=1}^{k+n}(r_{1i}^{\rm opt} - r_{1i}^{\rm Du})\xi_i > 0$, until $t_{k+n}$, $S^{\rm opt}$ has transmitted more bits to the stronger user than $S^{\rm Du}$ does; therefore, all the stronger user's bits arrived have not been transmitted by $S^{\rm Du}$ at the end of epoch $(k+n)$. Also, $r_{2(k+n)}^{\rm Du} = h_{2}(P_{k+n}^{\rm Du},r_{1(k+n)}^{\rm Du}) > h_2(P_{k+n}^{\rm opt},r_{1(k+n)}^{\rm opt}) \geq 0$. Hence neither of the two conditions in Lemma~\ref{lmm:ConstantRate1} holds and stronger user's rate cannot increase at $t_{k+n}$, which implies $r_{1k}^{\rm Du} \not< r_{1k}^{\rm opt}$. Thus, we are left with the case $r_{1k}^{\rm Du} > r_{1k}^{\rm opt}$. If $r_{1i}^{\rm opt} = r_{1k}^{\rm opt}, i \in \{k,...,s+m\}$, then $\sum_{i=1}^{s+m}(r_{1i}^{\rm Du} - r_{1i}^{\rm opt})\xi_i > 0$ , which contradicts the fact that DuOpt respects bit feasibility. Hence, stronger user's rate in $S^{\rm opt}$ cannot remain constant after epoch $k$. Then we should have $r_{1i}^{\rm opt} = r_{1k}^{\rm opt}, i \in \{k,...,k+n\}$ and $r_{1(k+n)}^{\rm opt} < r_{1(k+n+1)}^{\rm opt}$. Since there is an increase in the stronger user rate, at least one of the conditions in Lemma~\ref{lmm:ConstantRate1} should hold at $t_{k+n}$. However, we have $\sum_{i=1}^{k+n}(P_{i}^{\rm Du} - P_{i}^{\rm opt})\xi_i > 0$ and $\sum_{i=1}^{k+n}(r_{1i}^{\rm Du} - r_{1i}^{\rm opt})\xi_i > 0$, which tells us that neither one of the conditions in Lemma~\ref{lmm:ConstantRate1} holds, which implies that this final case is also not possible. Hence, we conclude that the case $P_{s}^{\rm Du} > P_{s}^{\rm opt}$ is not possible.

\item Now consider the case $P_{s}^{\rm Du} < P_{s}^{\rm opt}$. We will prove that this case is also not possible by following a similar method to the one in case (\ref{prf:item1}). First, suppose that the power of $S^{\rm Du}$ increases after $s^{th}$ epoch. This increase cannot be due to an energy constraint, since $S^{\rm opt}$ consumes more energy than $S^{\rm Du}$ does until the increase in power. Hence, it should be due to data constraint and under WUFBC both user data constraints should be active. That is, $S^{\rm Du}$ transmits all the feasible data until the increase in power. This implies that while consuming less energy, $S^{\rm Du}$ transmits at least the same number of bits than $S^{\rm opt}$ does, which contradicts the optimality of $S^{\rm opt}$. Thus, power of $S^{\rm Du}$ cannot increase after epoch $s$. Also, it cannot decrease in time, otherwise a local optimization results in either a gap or excess energy that propagates till the end of the schedule and transmission duration decreases. Therefore, we power of $S^{\rm Du}$ should stay constant after epoch $s$ until $T^{\rm opt}$. Now, we will analyze the rate assignments for both schedules. Let the transmission of $S^{\rm opt}$ end in epoch $(s+m)$ for $m \geq 0$ and suppose that $r_{1i}^{\rm Du} = r_{1i}^{\rm opt} \mbox{~~} \forall i < k, \mbox{~} 0 < k < (s+m)$. At the $k^{th}$ epoch there are three possible cases: $r_{1k}^{\rm Du} > r_{1k}^{\rm opt}$, $r_{1k}^{\rm Du} < r_{1k}^{\rm opt}$ and $r_{1k}^{\rm Du} = r_{1k}^{\rm opt}$. 
We will first consider the case $r_{1k}^{\rm Du} > r_{1k}^{\rm opt}$ and prove that this is not possible. Let $r_{1k}^{\rm Du} > r_{1k}^{\rm opt}$ and consider the rate of the stronger user in $S^{\rm opt}$ after $k^{th}$ epoch. It cannot stay constant until $T^{\rm opt}$, because it contradicts the fact that $S^{\rm opt}$ transmits all the feasible bits before $T^{\rm opt}$, \ie, $\sum_{i=1}^{s+m}(r_{1i}^{\rm Du} - r_{1i}^{\rm opt})\xi_i > 0$. Since the stronger user's rate in $S^{\rm opt}$ cannot decrease by Lemma~\ref{lmm:ConstantRate1}, it should increase at the end of epoch $(k+n)$ for $0 \leq n < (s+m-k)$, \ie, 
$r_{1(k+n)}^{\rm opt} < r_{1(k+n+1)}^{\rm opt}$. However, we have $\sum_{i=1}^{k+n}(r_{1i}^{\rm Du} - r_{1i}^{\rm opt})\xi_i > 0$ and $r_{2(k+n)}^{\rm opt} = h_{2}(P_{k+n}^{\rm opt},r_{1(k+n)}^{\rm opt}) > h_2(P_{k+n}^{\rm Du},r_{1(k+n)}^{\rm Du}) \geq 0$ which implies that none of the conditions in Lemma~\ref{lmm:ConstantRate1} holds and the stronger user's rate in $S^{\rm Du}$ cannot increase after epoch $k$. Hence, we conclude that $r_{1k}^{\rm Du} \not> r_{1k}^{\rm opt}$. Now we consider the case $r_{1k}^{\rm Du} < r_{1k}^{\rm opt}$. Suppose that the stronger user's rate in $S^{\rm opt}$ increase at epoch $(k+n)$ for $0 \leq n < (s+m-k)$. This increase in stronger user's rate requires that at least one of the conditions in Lemma~\ref{lmm:ConstantRate1} should hold. However, we have $\sum_{i=1}^{k+n}(r_{1i}^{\rm Du} - r_{1i}^{\rm opt})\xi_i > 0$ and $r_{2(k+n)}^{\rm opt} = h_{2}(P_{k+n}^{\rm opt},r_{1(k+n)}^{\rm opt}) > h_2(P_{k+n}^{\rm Du},r_{1(k+n)}^{\rm Du}) \geq 0$, so the stronger user's rate in $S^{\rm Du}$ cannot increase, \ie, $r_{1k}^{\rm Du} \not> r_{1k}^{\rm opt}$. Since the stronger user's rate in $S^{\rm opt}$ cannot decrease by Lemma~\ref{lmm:ConstantRate1}, it should stay constant until $T^{\rm opt}$. 

Thus far we have shown that if $S^{\rm Du}$ is different than $S^{\rm opt}$, then $S^{\rm Du}$ cannot have higher power level than $S^{\rm opt}$ until $T^{\rm opt}$. Moreover, if power level of $S^{\rm Du}$ becomes lower than that of $S^{\rm opt}$, then it should stay constant until $T^{\rm opt}$ and if the stronger user's rate of $S^{\rm Du}$ becomes lower than that of $S^{\rm opt}$, then it should stay constant until $T^{\rm opt}$. These results are shown in the general case in Fig.~\ref{fig:duopt_vs_optimal}.


Now, let $\tilde{b}_2^{\rm Du}$ and $\tilde{b}_2^{\rm opt}$ be the number of bits transmitted to the weaker user till $t_{s+m+l-1}$ by $S^{\rm Du}$ and till $T^{\rm opt}$ by $S^{\rm opt}$, respectively. Then, we have 
\small
\begin{eqnarray}
\tilde{b}_2^{\rm Du}-\tilde{b}_2^{\rm opt} &=&\sum_{i=1}^{s+m+l} \xi_i h_2(P_i^{\rm Du}, r_1^{\rm Du}) - \sum_{i=1}^{s+m+l} \xi_i h_2(P_i^{\rm Opt}, r_1^{\rm Opt})\nonumber\\
&=&\sum_{i=k}^{s-1} \xi_i \underbrace{(h_2(P_i^{\rm Du}, r_1^{\rm Du})-h_2(P_i^{\rm Opt}, r_1^{\rm Opt}))}_{>0} + \left(\sum_{i=s}^{s+m+l}\xi_i\right) h_2(P_s^{\rm Du}, r_{1s}^{\rm Du}) - \sum_{i=s}^{s+m} \xi_i h_2(P_i^{\rm Opt}, r_{1i}^{\rm opt})\nonumber\\
&>&\left(\sum_{i=s}^{s+m+l}\xi_i\right) h_2(P_s^{\rm Du}, r_{1s}^{\rm Du}) - \sum_{i=s}^{s+m} \xi_i h_2(P_i^{\rm Opt}, r_{1i}^{\rm opt})\nonumber\\
&\geq& \left(\sum_{i=s}^{s+m+l}\xi_i\right) h_2(\sum_{i=s}^{s+m}\frac{\xi_i}{\sum_{i=s}^{s+m+l}\xi_i}P_i^{\rm Opt}, r_{1s}^{\rm Du}) - \sum_{i=s}^{s+m} \xi_i h_2(P_i^{\rm Opt}, r_{1i}^{\rm opt})\nonumber\\
&>&\left(\sum_{i=s}^{s+m+l}\xi_i\right) \left(\sum_{i=s}^{s+m}\frac{\xi_i}{\sum_{i=s}^{s+m+l}\xi_i} h_2(P_i^{\rm Opt}, r_{1s}^{\rm Du})\right) - \sum_{i=s}^{s+m} \xi_i h_2(P_i^{\rm Opt}, r_{1i}^{\rm opt})\nonumber\\
&=&\sum_{i=s}^{s+m}\xi_i h_2(P_i^{\rm Opt}, r_{1s}^{\rm Du}) - \sum_{i=s}^{s+m} \xi_i h_2(P_i^{\rm Opt}, r_{1i}^{\rm opt})\nonumber\\
&=&\sum_{i=s}^{s+m}\xi_i \underbrace{(h_2(P_i^{\rm Opt}, r_{1s}^{\rm Du}) - h_2(P_i^{\rm Opt}, r_{1i}^{\rm opt})}_{>0}\nonumber\\
&>& 0 \label{eqn:BitFeasibility_5}
\end{eqnarray}
\normalsize

From (\ref{eqn:BitFeasibility_5}), $S^{\rm Du}$ transmits more bits to the weaker user than $S^{\rm opt}$ does, then this final case also cannot happen. Therefore, we conclude that the schedule returned by \emph{DuOpt} cannot be different than the unique optimal schedule, \ie, $S^{\rm Du} = S^{\rm opt}$. \enp

\end{enumerate}

\section{Numerical Example} 
\begin{figure}[hbt]
\centering \includegraphics[scale=0.5]{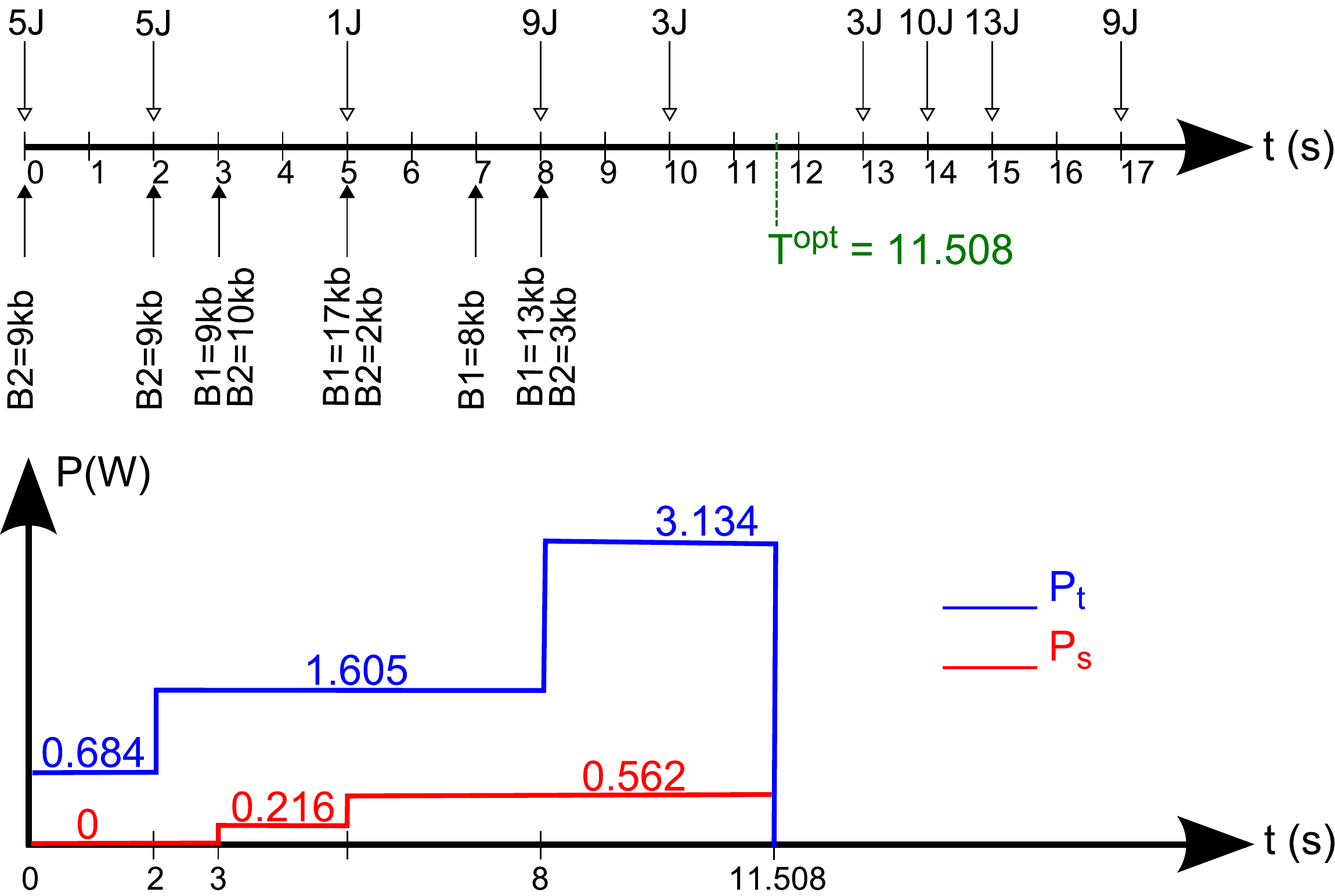}
\caption{(a) An illustration of energy harvest and bit arrival sequences. (b) Final schedule calculated by DuOpt, where $P_{s}$ is the power used for transmission to stronger user and $P_{t}$ is the total power used by the transmitter.}
\label{fig:NumEx}
\end{figure}

Consider a two-user AWGN BC with 1Khz bandwidth and $N_0=10^{-12}$ Watts/Hz. Path loss factors on the links of stronger and weaker user are assumed to be 70dB and 75dB, respectively. Amounts and instants of energy harvests and bit arrivals are depicted in Fig.~\ref{fig:NumEx}. Under these circumstances DuOpt algorithm is run and final schedule is calculated as drawn in Fig.~\ref{fig:NumEx}


\section{Conclusion} With the new advances in energy harvesting technologies, optimization of communication systems that depend on renewable energy resources has emerged as an important problem. A large body of recent research effort in the field has focused on transmission scheduling policies~\cite{MAAEUHE2010, YaOU2010, YaU2010, TuYe2010, Shroff2011,Tassiulas2010}. This paper  continued the work on to the solution of the previous formulation of the problem stated in~\cite{MAAEUHE2010} and~\cite{YaOU2010}.  In particular, it aimed to find the power and rate allocation policy in a Broadcast Channel with two users, that minimizes the total transmission completion time of data that becomes available at arbitrary points in time, with energy that is harvested at arbitrary points in time. It should be noted that this line of work relies on offline problem formulations where  exact knowledge of data and energy arrival events is assumed. The approach is useful for obtaining structures and guidelines as well as benchmarks for online problems which may be more applicable in practical transmision scenarios.

In this study, structural properties about the solution of the general optimal offline broadcast packet scheduling policy for an energy harvesting broadcast channel have been established. The uniqueness of the optimal policy under the weak-user-full-buffer condition (WUFBC) has been shown. An iterative algorithm, DuOpt, that returns a feasible schedule which possesses the same structural policies that the optimal was shown to have is devised. Moreover, DuOpt was shown to obtain the unique optimal policy under WUFBC. 

Proving the uniqueness of the optimal policy and optimality of DuOpt in general are two goals of ongoing work. The direction set for future work related to the problem presented in this paper also includes online problems. 

\section{Appendix}
\subsection{Proof of Lemma~\ref{lmm:PowernonDecreasing}}
\begin{figure}[htpb]
\centering \includegraphics[scale=0.35]{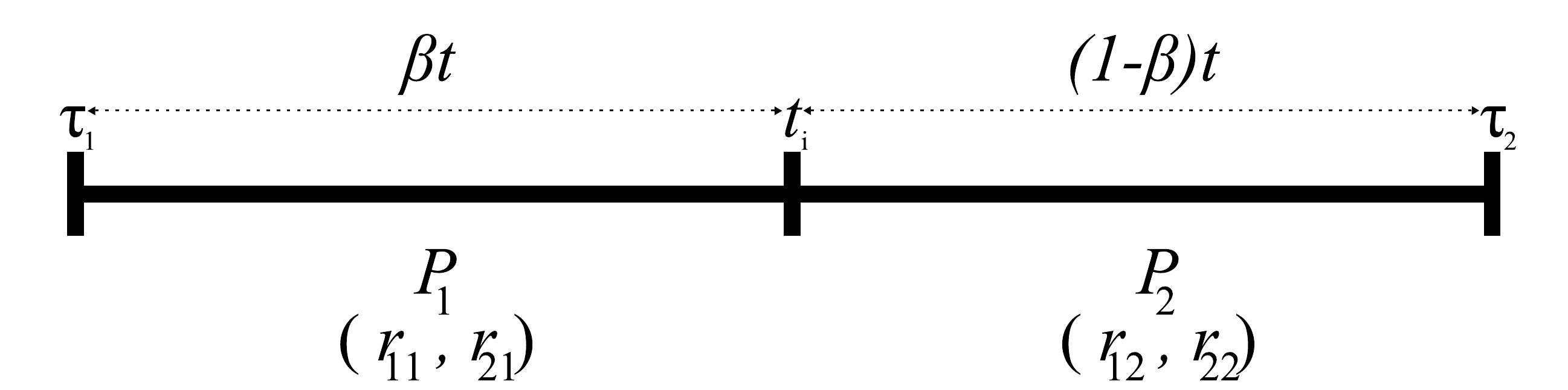}
\caption{Illustration of the transmission scheme used in proof of Lemma 2.}
\label{fig:CloserPowers}
\end{figure}

The claim is that power is non-decreasing with epoch number $i$. Equivalently power is non-decreasing in time. To show this, we will argue that given a schedule in which power decreases at some time $t_i$, this schedule can only be improved by equating the power levels before and after $t_i$. Consider a time interval $(\tau_1, \tau_2)$, so that power is constant at $P_1>0$ during $(\tau_1,t_i)$, and at $P_2<P_1$ during $(t_i,\tau_2)$. As illustrated in Fig~\ref{fig:CloserPowers}, let $t=\tau_2-\tau_1$, and the lengths of the constant-power slots be $\beta t$ and $(1-\beta)t$. Denote the rate pairs in the $1^{st}$ and $2^{nd}$ slots as ($r_{11}$, $r_{21}$) and ($r_{12}$, $r_{22}$), respectively. 
We will show that keeping the total consumed energy constant, and transferring some amount of energy $\Delta E$ from the first slot to the second such that power levels are reallocated closer to each other, the sender can transmit at least the same number of bits within the same duration.
Let us denote the average rate of the weaker user as 
$\bar{r_{2}} \triangleq \beta r_{21} + (1-\beta) r_{22}$. 
Provided $r_{21}>0$, the sender could transfer some energy and some of user 2's bits from the first epoch to the second while keeping user 1's rates ${r_{11}}$  and ${r_{12}}$ constant. As energy and bits are simply being deferred for later use, this operation does not violate feasibility.
Specifically, let 
\begin{equation}
P_1^{'}= P_1 - (1-\beta)\Delta P \mbox{~,~}P_2^{'}= P_2 + \beta \Delta P.
\label{eq:Del_P1}
\end{equation}
such that the new power allocation to the slots is $(P_1^{'},P_2^{'})$ satisfying $P_2 \leq P_2^{'} \leq P_1^{'} \leq P_1$. With this new allocation, the weaker user's rate in the first slot is $h_2(P_1^{'},{r_{11}})\leq h_2(P_1,{r_{11}})=r_{21}>0$. Its new new average rate over the duration of $t$ is:
\begin{eqnarray}
\bar{\bar{r_{2}}}&=& h_2(P_1^{'},{r_{11}})\beta + h_2(P_2^{'},{r_{12}})(1-\beta) \nonumber \\
&\geq & h_2(P_1,r_{11})\beta + h_2(P_2,r_{12})(1-\beta) \label{eq:Avg_2nd_User_Rate_1}\\
&=\bar{r_{2}} \nonumber
\end{eqnarray}
This is shown by straightforward application of the properties listed in section II.

\normalsize

In the remaining case which is $r_{21}=0$, we know that $r_{11}>0$ must hold (as $P_1>0$.) In this case, the allocation can similarly be improved by bringing power levels closer and transferring some of the first user's bits to the right, while keeping the rate allocation of the weaker user unchanged. Let $\bar{r_{1}} \triangleq \beta r_{11} + (1-\beta) r_{12}$ be the average rate of the stronger user over the duration $t$. After the reallocation, the average rate of the stronger user becomes
\begin{eqnarray}
\bar{\bar{r_{1}}}&=& h_1(P_1^{'},{r_{21}})\beta + h_1(P_2^{'},{r_{22}})(1-\beta) \nonumber \\
&\geq & h_1(P_1,r_{21})\beta + h_1(P_2,r_{22})(1-\beta) \label{eq:Avg_1st_User_Rate_1}\\
&=& \bar{r_{1}} \nonumber
\end{eqnarray}

\eqref{eq:Avg_1st_User_Rate_1} follows from the fact that
\begin{eqnarray}
&&h_1(P_1^{'},{r_{21}})\beta + h_1(P_2^{'},{r_{22}})(1-\beta) - h_1(P_1,r_{21})\beta - h_1(P_2,r_{22})(1-\beta)\geq 0
\label{eq:f2}
\end{eqnarray}
for all $\beta=\{0,1\}$ with equality achieved at $\beta=0,1$. 
\begin{eqnarray}
q(\beta)&=&h_1(P_1-(1-\beta)\Delta P,{r_{21}})\beta + h_1(P_2+\beta \Delta P,{r_{22}})(1-\beta) \nonumber \\
&&- h_1(P_1,r_{21})\beta - h_1(P_2,r_{22})(1-\beta).
\end{eqnarray}
\normalsize
We can show that \eqref{eq:f2} holds by proving $q(\beta)$ is concave in $\beta$.   

The $1^{st}$ and $2^{nd}$ order derivatives of $q$ with respect to $\beta$ are  the following\footnote{$h_{1_{x}}$ and $h_{1_{xx}}$ represent the first and second order partial derivatives of $h_{1}$ with respect to $P$, respectively.}

\begin{eqnarray}
\frac{\partial q}{\partial \beta} &=& h_1(P_1-(1-\beta)\Delta P,{r_{21}})+\beta \left\{ h_{1_{x}}(P_1-(1-\beta)\Delta P,{r_{21}})(\Delta P)\right\}\nonumber \\
&& -h_1(P_2+\beta \Delta P,{r_{22}}) + (1-\beta) \left\{ h_{1_{x}}(P_2+\beta \Delta P,{r_{22}}) (\Delta P)\right\}\nonumber \\
&&- h_1(P_1,r_{21}) + h_2(P_2,r_{22})
\end{eqnarray}
\begin{eqnarray}
\frac{\partial ^2 q}{\partial \beta^2} &=&2(\underbrace{h_{1_{x}}(P_1-(1-\beta)\Delta P,{r_{21}})(\Delta P)-h_{1_{x}}(P_2+\beta \Delta P,{r_{22}}) (\Delta P)}_{\leq 0})\nonumber\\
&&+\beta \left\{ \underbrace{h_{1_{xx}}(P_1-(1-\beta)\Delta P,{r_{21}})(\Delta P)^2}_{\leq 0}\right\}\nonumber\\
&&+(1-\beta)\left\{ \underbrace{h_{1_{xx}}(P_2+\beta \Delta P,{r_{22}})(\Delta P)^2}_{\leq 0}\right\}\nonumber \\
&&\leq 0\label{eq:qxx} 
\end{eqnarray}

According to the properties listed in section II, ~\eqref{eq:qxx} always holds if $r_{21} \geq r_{22}$. Hence $q$ is concave in $\beta$, if $r_{21}=0$.
We conclude that a policy that contains a drop in power level is sub-optimal. \enp 

\subsection{Proof of Lemma~\ref{lmm:CloserPowers}}

To reach contradiction, suppose that power increases at time $t_i$ ($P_{i+1}>P_{i}$). We will show that if none of the conditions~\eqref{cnd:a},~\eqref{cnd:b} or~\eqref{cnd:c} hold, then it is possible to improve the schedule by transferring some energy from the $(i+1)^{th}$ epoch to the $i^{th}$. Assuming the $i^{th}$ epoch length is $\beta t$ and the $(i+1)^{th}$ is $(1-\beta) t$, after bringing power levels closer,
\begin{equation}
P_i^{'}= P_i +\beta\Delta P \mbox{~,~}P_{i+1}^{'}= P_{i+1} - (1-\beta)\Delta P.
\label{eq:Del_P2}
\end{equation} 
Observe that if we treat $P_i$ as $P_2$ and $P_{i+1}$ as $P_1$, then~\eqref{eq:Del_P2} becomes identical with~\eqref{eq:Del_P1}. This implies that at least the same number of bits could be transmitted to weaker user, if we can bring power levels closer while keeping the stronger user's rates constant. In addition to this, the allocation could also be improved by bringing power levels while keeping the weaker user's rates constant, in case $r_{2(i)} \geq r_{2(i+1)}$. Consequently, equations~\eqref{eq:Avg_2nd_User_Rate_1} and~\eqref{eq:Avg_1st_User_Rate_1} hold. 

It is straightforward that we cannot bring power levels any closer when condition~\eqref{cnd:a} holds, due to the energy causality constraint. Secondly, it also doesn't yield a better schedule, if we can not transfer data from the latter epoch to the former(condition~\eqref{cnd:b}). When it is possible to transfer some positive amount of energy from the $(i+1)^{th}$ epoch to the $i^{th}$, as shown in~\eqref{eq:Avg_2nd_User_Rate_1}, we can always improve allocation while keeping the rates of the stronger user the same. Although bringing power levels closer while keeping the rates of the stronger user the same is not feasible in case weaker user's data constraint is active, we may still improve allocation as proved in~\eqref{eq:Avg_1st_User_Rate_1}. Nevertheless, this time we require $r_{2(i)} \geq r_{2(i+1)}$. As rate of the weaker user can only rise upon a data arrival for the weaker user in case weaker user's data constraint is active, condition~\eqref{cnd:c} describes the last case that we may not improve allocation by bringing power levels closer. 

We have thus shown that this set of three conditions contains all the cases in which power can rise, if none of these hold, then power cannot rise. It is straightforward to show that this set cannot be further reduced by finding counterexamples for the claim that if any one of~\eqref{cnd:a},~\eqref{cnd:b} or~\eqref{cnd:c} is satisfied, then power may rise at time $t_i$. \enp

\subsection{Proof of Corollary} \begin{enumerate}
\item Suppose that power increases upon a bit arrival for the weaker user occurring at $t_i$. As there is no energy constraint at $t_i$, bringing power levels closer does not contradict with the energy causality in this case. This implies that conditions~\eqref{cnd:b} or~\eqref{cnd:c} stated in Lemma~\ref{lmm:CloserPowers} must hold. However, we know that there is a data arrival for the weaker user at $t_i$, so if~\eqref{cnd:b} were true, then~\eqref{cnd:c} would be true as well. Therefore, condition~\eqref{cnd:c} holds in either case. \enp
\item Suppose that power increases upon a bit arrival for the first user. With similar reasoning to part-1, condition~\eqref{cnd:a} of Lemma~\ref{lmm:CloserPowers} cannot hold. As there is also no data arrival for the weaker user, condition~\eqref{cnd:b} must be satisfied. \enp
\item As there is no data arrival at the time when power increases, the only possibility that power increases upon an energy harvest is condition~\eqref{cnd:a} of Lemma~\ref{lmm:CloserPowers}. \enp
\end{enumerate}

\subsection{Proof of Lemma~\ref{lmm:ConstantRate1}} Suppose that $r_{1i} \leq r_{1(i+1)}$. One can find a better schedule by bringing the rates of the stronger user closer by Lemma 6 of~\cite{MAAEUHE2010}\footnote{Lemma~6 of~\cite{MAAEUHE2010} originates from the observation in~\cite{YaOU2010} (See Lemma~4 and Corollary~1) that there is a cut-off level for the total power, below which no power is assigned to the weaker user.}. Therefore, the stronger user's rate cannot decrease. However, the stronger user's rate may increase because it may be against to either bit or energy causality to transfer some stronger user bits from epoch $i+1$ to $i$. Firstly, it is against \emph{bit causality} to transfer some stronger user bits from epoch $i+1$ to epoch $i$, if the first condition holds. Secondly, if the second condition is satisfied we cannot bring stronger users rates closer to each other as it would violate \emph{energy causality}. \enp\\

\subsection{Details of the Local Optimization}
\label{app:local_optimization}

\begin{figure}[htpb]
\begin{center}
\includegraphics[scale=0.8]{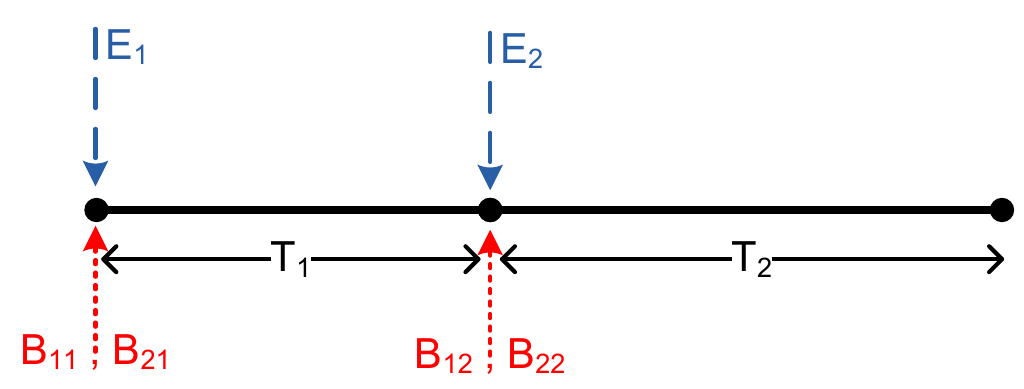} 
\caption{Illustration of the local optimization problem with two epochs. $B_{ij}$ represents the data arrival for the $j^{th}$ user at the beginning of the $i^{th}$ epoch. Similarly, $E_i$ represents the energy harvest at the beginning of the $i^{th}$ epoch.}
\label{fig:Local_Constraints}
\end{center} 
\end{figure}

Consider the local optimization problem given in Fig.~\ref{fig:Local_Constraints}, where $B_{ij}$ is the data arrival for the $i^{th}$ user, $E_j$ is the energy harvest at the beginning of $j^{th}$ epoch for $i,j \in \{1,2\}$. $T_1$ is the length of the first epoch and $T_2$ is the transmit duration in the second epoch. Let $P_{ij}$ and $r_{ij}$ be the power and rate assigned to the $i^{th}$ user during $j^{th}$ epoch after optimization. The energy and data causality constraints for the local optimization problem are as follows:
\begin{eqnarray*}
\label{eq:energy_constraint} E_1 &\geq& P_{11} \cdot T_1 \\
\label{eq:user1data_constraint} B_{11} &\geq& r_{11} \cdot T_1 \\
\label{eq:user2data_constraint} B_{21} &\geq& r_{21} \cdot T_1.
\end{eqnarray*}

The structure of the solution changes if either one of the constraints satisfied with equality. Since there are $3$ different constraints, after optimization one will encounter one of the $2^3 = 8$ results. We have studied all $8$ cases and derived solutions to each one of them for both energy minimization and time minimization functions. In each case, the solution can be calculated analytically for energy minimization or it can be found iteratively for time minimization functions. Before starting the optimization, if one already knows which constraints should be satisfied with equality, then the result could be obtained solving just that case. Otherwise, one should compute the results for each case and then select the best one\footnote{Best result is the one that consumes minimum energy for the \emph{energy minimization} function and one that has the minimum transmit time for the \emph{time minimization} function.} that respects energy and bit causalities. Next, we will analyze one of the cases and present the algorithms for both energy and time minimization functions. The analyzes of the remaining cases can be done in a similar fashion~\cite{Hakan_Thesis}.

\begin{figure}[htpb]
\begin{center}
\includegraphics[scale=0.65]{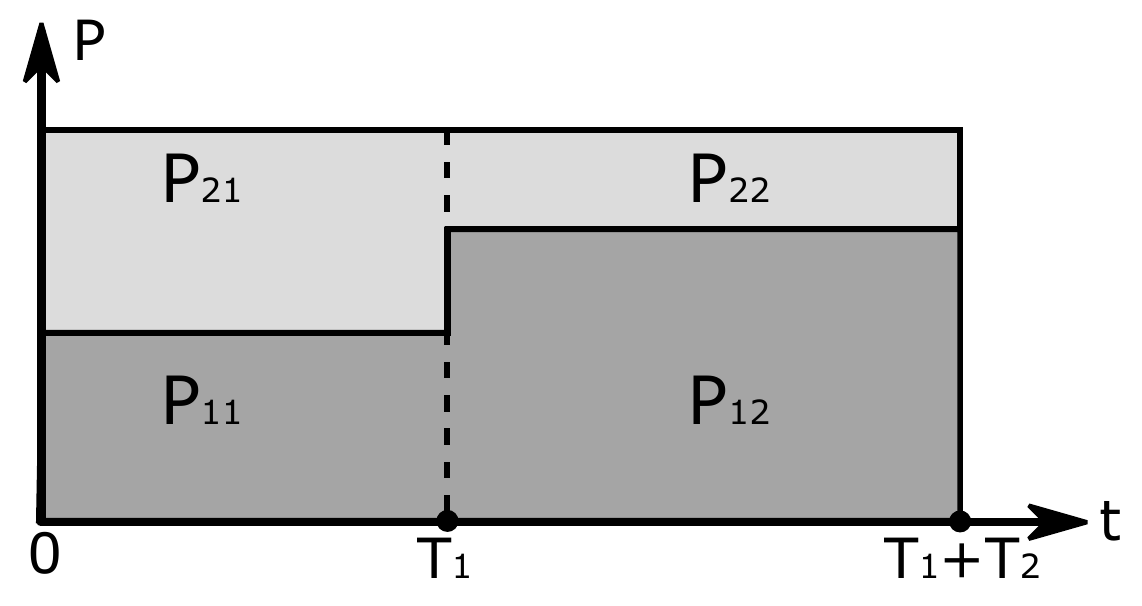} 
\caption{Optimal power allocation for the stronger and the weaker user if only the stronger user data causality constraint is active. Dark shaded levels represent the stronger user's power levels, whereas the light shaded ones represent the weaker user's power levels. The total transmit power stays constant.}
\label{fig:LocalOpt_B1}
\end{center} 
\end{figure}

\paragraph*{Local Optimization when only stronger user data constraint is active}
One of the possible structures of the local optimal solution is when only the stronger user data constraint is active. In this case, total transmit power level in local optimal solution should be constant (See Lemma~\ref{lmm:CloserPowers}). Under this condition, a possible illustration of optimal power allocation is depicted in Fig.~\ref{fig:LocalOpt_B1}. Since the total power should stay constant during transmission, we should have $P_{11} + P_{21} = P_{12} + P_{22}$. Moreover, we should have $P_{11} \leq P_{12}$ since only the stronger user bit causality is met.


For a given transmission completion time $T_1 + T_2$, the optimum schedule that minimize energy consumption can be found as follows:
\begin{eqnarray}
\label{eq:power_opt_b1} P_{11} &=& \frac{\sigma^2}{s_1}(2^{\frac{2B_{11}}{T_1}} -1) \mbox{~~,~~}
P_{12} = \frac{\sigma^2}{s_1}(2^{\frac{2B_{12}}{T_2}} -1) \mbox{~~and~~}
P_{22} = P_{11} + P_{21} - P_{12} \\
\label{eq:b2_opt} B_2 &=& B_{21} + B_{22} =  \frac{T_1}{2} \log_2(1+\frac{P_{21} s_2}{P_{11}s_2 + \sigma^2}) + \frac{T_2}{2} \log_2(1 + \frac{P_{22} s_2}{P_{12} s_2 + \sigma^2}).
\end{eqnarray}
From~\eqref{eq:power_opt_b1} and~\eqref{eq:b2_opt} we derive
\begin{equation}
P_{21} = -P_{11} + \frac{1}{s_2}\left[\left(2^{2B_2}(P_{11}s_2+\sigma^2)^{T_1}(P_{12}s_2+\sigma^2)^{T_2} \right)^{\frac{1}{T_1 + T_2}} -\sigma^2\right].
\end{equation}
Then, total energy consumed in two epochs is calculated by
\begin{equation}
\label{eq:min_energy}E_{min} = (P_{11} + P_{21})(T_1 + T_2).
\end{equation}

Minimum energy to transmit $B_1 = B_{11} + B_{12}$ and $B_2= B_{21} + B_{22}$ bits to the users in two epochs, $E_{min}$, can be calculated by setting $T_2$ as the length of the second epoch in (\ref{eq:min_energy}).

Substituting \eqref{eq:power_opt_b1} into \eqref{eq:b2_opt} and arranging terms we obtain
\begin{eqnarray}
B_2 &=& \frac{T_1}{2}\log_2\left( \frac{\frac{E_1+E_2}{T_1+T_2} + \frac{\sigma^2}{s_2}} {\frac{\sigma^2}{s_1}(2^{\frac{2B_{11}}{T_1}}-1)+\frac{\sigma^2}{s_2}}\right) + \frac{T_2}{2}\log_2\left( \frac{\frac{E_1+E_2}{T_1+T_2} + \frac{\sigma^2}{s_2}} {\frac{\sigma^2}{s_1}(2^{\frac{2B_{12}}{T_2}}-1)+\frac{\sigma^2}{s_2}}\right) \nonumber \\
&=& \frac{T_1+T_2}{2} \log_2\left(\frac{E_1+E_2}{T_1+T_2} + \frac{\sigma^2}{s_2}\right) \nonumber \\ &&- \frac{T_1}{2}\log_2\left(\frac{\sigma^2}{s_1}(2^{\frac{2B_{11}}{T_1}} -1) + \frac{\sigma^2}{s_2}\right) - \frac{T_2}{2}\log_2\left(\frac{\sigma^2}{s_1}(2^{\frac{2B_{12}}{T_2}} -1) + \frac{\sigma^2}{s_2}\right) \nonumber
\end{eqnarray}

The first and second order derivatives of $B_2$ with respect to $T_2$ are as follows:

\begin{eqnarray}
\frac{\partial B_2}{\partial T_2} &=& \frac{1}{2}\log_2\left(\frac{E_1+E_2}{T_1+T_2} + \frac{\sigma^2}{s_2}\right) - \frac{1}{2\ln(2) \left(1 + \frac{\sigma^2}{s_2}\frac{T_1+T_2}{E_1+E_2}\right)} \nonumber \\
&& -\frac{1}{2}\log_2\left(\frac{\sigma^2}{s_1}2^{\frac{2B_{12}}{T_2}}+\sigma^2\frac{s_1-s_2}{s_1 s_2}\right) + \frac{B_{12}}{T_2} \frac{1}{1 + \frac{s_1-s_2}{s_2}2^{-\frac{2B_{12}}{T_2}}} \label{B2_derivative1} \nonumber \\
\frac{\partial^2 B_2}{\partial T_{2}^{2}} &=& -\frac{1}{2\ln(2)(T_1+T_2)\left(1+ \frac{\sigma^2}{s_2}\frac{T_1+T_2}{E_1+ E_2}\right)^2} - \frac{2\ln(2)B_{12}^2 \frac{s_1-s_2}{s_2}2^{-\frac{2B_{12}}{T_2}}}{T_2^3\left(1+\frac{s_1-s_2}{s_2}2^{-\frac{2B_{12}}{T_2}}\right)^2} \label{B2_derivative2} \\
&<& 0  \nonumber
\end{eqnarray}

As shown in~(\ref{B2_derivative2}), second derivative of $B_2$ is always negative for $s_1 > s_2$, which implies that $B_2$ is a strictly concave function of $T_2$. As $T_2$ goes to infinity, $B_2$ is as follows, 

\begin{eqnarray}
\lim_{T_2\to\infty} B_2 &=& -\frac{T_1}{2}\log_2\left(\frac{s_2}{s_1}(2^{\frac{2B_{11}}{T_1}} -1) +1\right)- \frac{s_2}{s_1} B_{12} + \frac{(E_1+E_2)s_2}{2\ln(2)\sigma^2} \label{B2_at_infinity} \nonumber \\
&>& -\infty \nonumber
\end{eqnarray}

Since $B_2$ is a strictly concave function of $T_2$ and goes to a finite number as $T_2$ goes to infinity, 
$B_2$ is an increasing strictly concave function of $T_2$ and there is a unique $B_2$ for each value of $T_2$.

In time minimization we have $E_1$, $E_2$, $T_1$, $B_2= B_{21} + B_{22}$, $B_{21}$, $B_{22}$; $s_1$ and $s_2$ are constant terms and $T_2$ is transmission time within the second epoch which is to be minimized. 
Since $B_2$ is an increasing concave function of $T_2$, we iteratively find $T_2$ that sends exactly $B_2$ bits to weaker user by using bisection method. Minimum time to transmit $B_1 = B_{11} + B_{12}$ and $B_2$ bits to the users in these two epochs can be calculated by $T_{min}^{(2)} = T_1+T_2$. Algorithm~\ref{alg:tmin_b1} presents a pseudo-code of time minimization algorithm for the case that only the stronger user bit causality event occurs.\\

\begin{algorithm}
\caption{Local time minimization algorithm when only the stronger user data constraint is active }
\label{alg:tmin_b1}
\begin{algorithmic}[1]
	\STATE {$T_{max} \gets T_2$, $T_{min} \gets 0$}
	\REPEAT
	\STATE $\hat{T} \gets (T_{max} + T_{min}) / 2$
	\STATE $\hat{B_2} \gets \frac{T_1}{2}\log_2\left( \frac{\frac{E_1+E_2}{T_1+\hat{T}} + \frac{\sigma^2}{s_2}} {\frac{\sigma^2}{s_1}(2^{\frac{2B_{11}}{T_1}}-1)+\frac{\sigma^2}{s_2}}\right) + \frac{\hat{T}}{2}\log_2\left( \frac{\frac{E_1+E_2}{T_1+\hat{T}} + \frac{\sigma^2}{s_2}} {\frac{\sigma^2}{s_1}(2^{\frac{2B_{12}}{\hat{T}}}-1)+\frac{\sigma^2}{s_2}}\right)$
	\IF{$\hat{B_2} < B_2$}
	\STATE $T_{min} \gets \hat{T}$
	\ELSE
	\STATE $T_{max} \gets \hat{T}$
	\ENDIF
	\UNTIL {$\hat{B_2} == B_2$}
	\STATE $T_{min}^{(2)} \gets \hat{T} + T_1 $
\end{algorithmic}
\end{algorithm}

\singlespacing

\newpage
\normalsize

\vspace{0.3in}

\vspace{0.3in}

\begin{figure}[htpb]
\begin{center}
\includegraphics[scale=0.55]{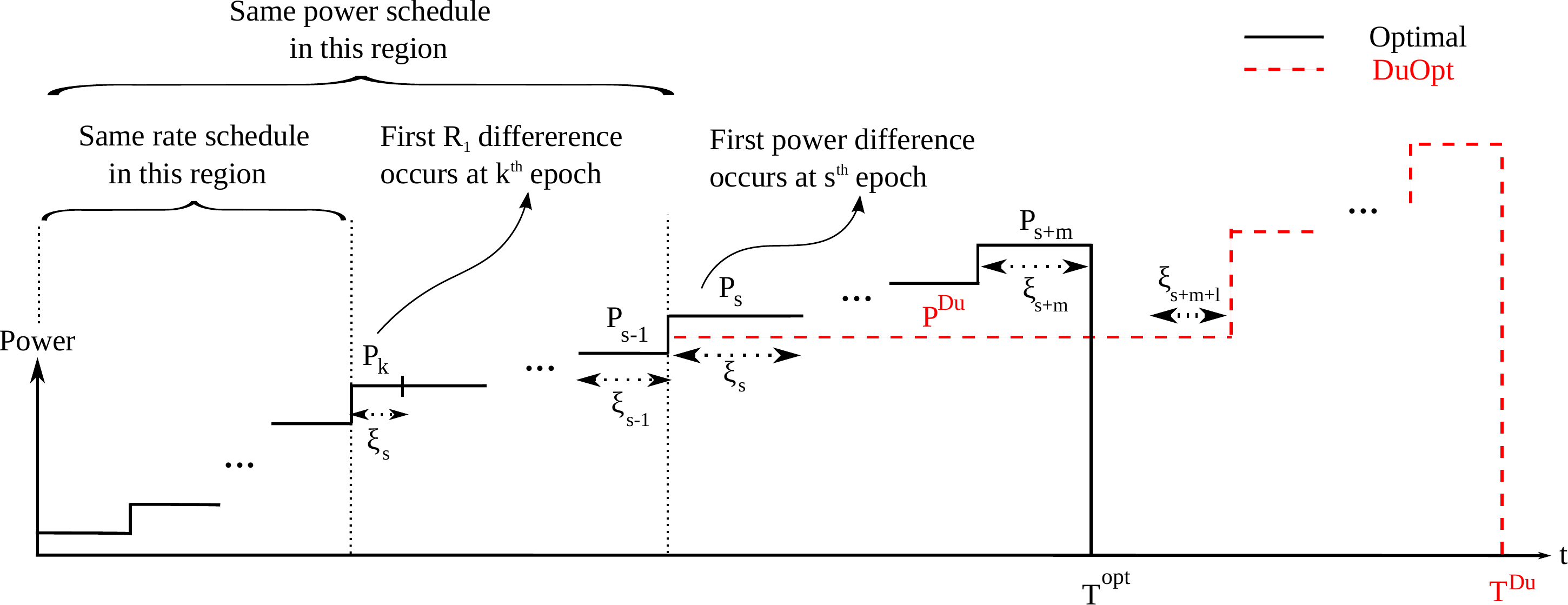} 
\caption{Illustration of the final case in the proof of Theorem~\ref{thm:DuOpt_optimal}.}
\label{fig:duopt_vs_optimal}
\end{center} 
\end{figure}
\vspace{0.3in}

\vspace{0.3in}

\vspace{0.3in}

\begin{thebibliography}{10}
\bibitem{OzErU} F. M. Ozcelik, H. Erkal and E. Uysal-Biyikoglu, ``Optimal Offline Packet Scheduling on an Energy Harvesting Broadcast Link,'' \emph{2011 IEEE Int. Symposium on Information Theory}, pp.2886-2890, Saint- Petersburg, Russia, Aug. 2011.
\bibitem{UEP02} E. Uysal-Biyikoglu and A. El Gamal, ``Energy-efficient Packet Transmission Over a Multi-access Channel,''  \emph{Proc. IEEE Intl. Symposium on Information Theory}, p.153, July 2002.
\bibitem{BeGa02} R. A. Berry and R. G. Gallager, ``Communication over fading channels with delay constraints,'' \emph{IEEE Transactions on Information Theory}, vol.48, pp.1135-1149, May 2002.
\bibitem{NuSr02} P. Nuggehalli, V. Srinivashan, and R. R. Rao, ``Delay constrained energy efficient transmission strategies for wireless devices,'' \emph{in Proc.IEEE INFOCOM}, vol.3, pp.1765-1772, New York, June 2002.
\bibitem{ZaMo09} M. A. Zafer and E. Modiano, ``A calculus approach to energy-efficient data transmission with quality of service constraints,'' \emph{IEEE/ACM Transactions on Networking}, vol.17, pp.898-911, June 2009.
\bibitem{YaU2010} J. Yang and S. Ulukus, ``Optimal Packet Scheduling in an Energy Harvesting Communication System,'' arXiv:1010.1295v1, 2010.
\bibitem{TuYe2010} K. Tutuncuoglu and A. Yener, ``Optimum transmission policies for battery limited energy harvesting nodes,'' arXiv:1010.6280], 2010.
\bibitem{OzTu2010} O. Ozel, K. Tutuncuoglu, J. Yang, S. Ulukus and A. Yener, ``Transmission with energy harvesting nodes in fading wireless channels: Optimal policies,'' \emph{Selected Areas in Communications, IEEE Journal on}, vol.29, no.8, pp.1732-1743, September 2011
\bibitem{MAAEUHE2010} M. A. Antepli, E. Uysal-Biyikoglu, and H. Erkal, ``Optimal Packet Scheduling on an Energy Harvesting Broadcast Link,'' \emph{Selected Areas in Communications, IEEE Journal on}, vol.29, no.8, pp.1721-1731, September 2011.
\bibitem{YaOU2010} J. Yang, O. Ozel and S. Ulukus, ``Broadcasting with an Energy Harvesting Rechargeable Transmitter,'' arXiv:1010.2993v1, 2010.
\bibitem{OzYa2010} O. Ozel, J. Yang and S. Ulukus, ``Broadcasting with a Battery Limited Energy Harvesting Rechargeable Transmitter,'' \emph{Modeling and Optimization in Mobile, Ad Hoc and Wireless Networks (WiOpt), 2011 International Symposium on}, pp.205-212, Princeton, NJ, May 2011.
\bibitem{Shroff2011} S. Chen, P. Sinha, N. B. Shroff and C. Joo, ``Finite-Horizon Energy Allocation and Routing Scheme in Rechargeable Sensor Networks,'' \emph{IEEE INFOCOM'11}, Shanghai China, April 2011.
\bibitem{Tassiulas2010} M. Gatzianas, L. Georgiadis and L. Tassiulas, ``Control of Wireless Networks with Rechargeable Batteries," in \emph{Trans. on Wireless Communications}, vol. 9, pp. 581-593, February 2010.
\bibitem{OzErU2011} F. M. Ozcelik, H. Erkal and E. Uysal-Biyikoglu, ``Optimal Offline Packet Scheduling on an Energy Harvesting Broadcast Link,'' \emph{IEEE International Symposium on Information Theory (ISIT)}, pp.2886-2890, St. Petersburg, Russia, July 2011.
\bibitem{Eu04} E. Uysal-Biyikoglu and A. El Gamal, ``On adaptive transmission for energy efficiency in wireless data networks,'' \emph{IEEE Transsactions on Information Theory}, vol.50, pp.3081-3094, Dec 2004.
\bibitem{UPE02} E. Uysal-Biyikoglu, B. Prabhakar, and A. El Gamal, ``Energy-efficient Packet Transmission over a Wireless Link,'' in \emph{IEEE/ACM Trans. Networking}, vol.10, pp.487-499, Aug. 2002. 
\bibitem{Zang69} W.I. Zangwill, {\em Nonlinear Programming}, Prentice-Hall, Engelwood Cliffs, New Jersey, Prentice-Hall, 1969.
\bibitem{Hakan_Thesis} H. Erkal, F. M. Ozcelik and E. Uysal-Biyikoglu ,{\em Optimal Offline Broadcasting with an Energy Harvesting Capable Transmitter}, Technical Report, METU, Ankara, Turkey, Available at http://www.eee.metu.edu.tr/$\scriptstyle\mathtt{\sim}$cng.
\end{thebibliography}
\end{document}